\documentstyle[prl,twocolumn,aps]{revtex}

\def\be{\begin{equation}}
\def\ee{\end{equation}}
\def\bex{\begin{eqnarray}}
\def\eex{\end{eqnarray}}
\def\mbar{\overline M}
\def\gbar{\overline G}

\begin{document}

\title{MR spectroscopy with parabolic magnetic field: spin-oscillator
coupling effect}

\author{Czes\l{}aw J. Lewa$^{1,}$\cite{poczta1},
Pawe\l{} Horodecki$^{2,}$\cite{poczta2},
Ryszard Horodecki$^{3,}$\cite{poczta3},
Micha\l{} Horodecki$^{3,}$\cite{poczta4}
}
\address{$^1$ Institute of Experimental Physics,
University of Gda\'nsk, 80--952 Gda\'nsk, Poland\\
$^2$Faculty of Applied Physics and Mathematics,
Technical University of Gda\'nsk, 80--952 Gda\'nsk, Poland\\
$^3$ Institute of Theoretical  Physics and Astrophysics,
University of Gda\'nsk, 80--952 Gda\'nsk, Poland\\
}

\maketitle

\begin{abstract}
The spectrum of the spin particle in oscillatory potential
subjected to external parabolic magnetic
field ${\bf B}=(B_0+Gx+\gbar x^2){\bf \hat z}$
is obtained. The structure of energy levels of the considered system 
allows to identify
the frequency of the oscillator via the spectrum of spin sublevels coming only
from {\it one} oscillatory level. The effect is due to the gradient terms in
the form of the field.
\end{abstract}

In a recent decade one can observe a rapid grow of interest in 
the Stern-Gerlach (SG) interaction \cite{Stern} within MR spectroscopy 
\cite{ency,Sidles91,Sidles92,Moore,Rugar,Lewa96a,Lewa96b}.
The main reasons  are the following:
(i) partial saturation of the possibilities of the conventional MR methods based on
Bloch-Purcell paradigm \cite{ency} (low sensitivity, especially for the
nuclear MR (NMR), growing costs of aparature, 
existence of
cheaper, competitive methods like e.g. ultrasonic ones), (ii) recent
reports on new and promising effects (e.g. atomic force microscopy
\cite{Sidles91,Sidles92}, MR
microimaging \cite{Moore,Rugar}, MR spectroscopy for selected Zeeman states
\cite{Lewa96a,Lewa96b} (iii) the need for new solutions in the domain of
NMR
quantum computing (the main problem in the recent implementations of
quantum computing is low signal intensity) \cite{NMRQC}
(iv) the recent development in the domain of ferromagnetics and
superconductors allowing to produce strong magnetic fields of required shapes.
In this situation one is seeking for new, both theoretical and experimental
solutions for MR, in particular by exploiting the SG interaction.


Recently Sidles \cite{Sidles92} has proposed the interesting
model of spin-$\frac{1}{2}$ particle mounted to the
harmonic oscillator. The particle is assumed to interact
with the magnetic field via Stern-Gerlach Hamiltonian plus some
time-dependent, constant in space magnetic field.
He solved the model and showed that
the wavepacket of the particle exhibits
the splitting into two parts moving
in opposite directions.
He also predicted the importance of
possible oscillator based NMR spectroscopy
for molecular imaging devices.

In this Letter we consider the coupled spin-oscillator system
in external heterogeneous magnetic field.
The model is solved exactly 
and provides  physical spectrum
in which the oscillator frequency is coupled to the
field derivatives parameters.
It implies that in principle one can  recognise the
frequency of oscillator basing only on
signals coming form transitions
between sublevels of {\it one} oscillatory level.

Let us consider the particle 
that subjects to the simple dynamic
due to the harmonic oscillator Hamiltonian
${\cal H}_{osc}^{\Omega}(x-a)$. Here $a$ is the position of the minimum of the
oscillator potential
with respect to the origin of the reference frame (we imagine that the particle
is in a sample of finite dimension $l$ situated in the origin of the
reference frame, so that $a$ can vary within the region $(-l,l)$).
The  Hamiltonian of the particle
\begin{equation}
{\cal H}_{osc}^{\Omega}(x)=\frac{-\hbar^{2}}{2m}\frac{d^{2}}{dx^{2}}+\frac{m\Omega^{2}x^{2}}{2}.
\end{equation}
has the following spectrum and corresponding eigenvectors:
\begin{eqnarray}
&&{\cal E}_{n}= \hbar \Omega( n + \frac{1}{2}) \nonumber \\
&&\Psi_{n}(\Omega;x)= (\frac{m\Omega}{ 2^{2n}(n!)^{2} \hbar \pi})^{\frac{1}{4}}
\cdot H_{n}(\sqrt{\frac{m\Omega}{\hbar}}x) \cdot exp(-\frac{m\Omega}{2\hbar}x^{2})
\label{widosc}
\end{eqnarray}
with  Hermite polynomials $H_{n}$.

Suppose now that the particle has the spin $S$ and
recall that the quantum operator corresponding
to the projection of the spin onto the $z$ axis is described as
$\hat{\mbox{\bf I}}=\sum_{M=-S}^{S} \hbar M |M \rangle \langle M|$
with the corresponding spectrum and eigenvectors
$\hbar M, |M \rangle $.

Consider the inhomogeneous magnetic field
\begin{equation}
\mbox{\bf B} (x)\equiv B(x)\hat{\mbox{\bf z}}=
(B_{0} + Gx+ \bar{G}x^{2})\hat{\mbox{\bf z}}
\label{pole}
\end{equation}
along $\hat{\mbox{\bf z}}$ axis.
Here $G$, $\bar{G}$ stands for the gradient
and second derivative parameters.
Note that the field has the value $ \mbox{\bf B}_{0}=
B_{0}\hat{\mbox{\bf z}}$ for $x=0$
an that in general it depends {\it quadratically}
on the spatial coordinate $x$.

Now if we put our particle in the field (\ref{pole})
then its Hamiltonian becomes
\begin{equation}
{\cal H}={\cal H}_{osc}^{\Omega}(x - a) \otimes I - \gamma  B(x)
\otimes \hat{\mbox{\bf z}} \hat{\mbox{\bf I}}
\end{equation}
In the above formula the difference
between spatial and spin degrees of freedom
has been stressed. In particular the
identity operator $I$ has been used.
The new energy spectrum of our particle
can be found seeking the eigenvectors in the
form $|\phi \rangle \otimes |M \rangle $ ($M= -S, -S+1, ..., S$). This  leads
effectively to the sequence of $2S+1$ $M$-dependent shifted
quantum oscillators.
Eigenvalues each of those oscillators can be easily solved
and the final energy spectrum of the particle
is given by
\begin{eqnarray}
&&E_{M,n}=\hbar\sqrt{\Omega^{2} - \frac{2\gamma\bar{G} \hbar M}{m}}(n+\frac{1}{2})
- \gamma(B_{0} + Ga+ \bar{G}a^{2})\hbar M  \nonumber \\
&& - \frac{\gamma^{2} (G + 2 {\bar G} a)^{2} {\hbar}^{2} M^{2}
}{2m(\Omega^{2} -\frac{2\gamma\bar{G}\hbar M}{m})} \nonumber \\
\label{WID}
\end{eqnarray}
and the corresponding eigenvectors can be written in the form
\begin{eqnarray}
\Phi_{M,n}= |\phi_{M,n}  \rangle \otimes |M \rangle
\end{eqnarray}
The spatial coordinate function is defined by
the oscillator eigenvectors (\ref{widosc})
in the following way
\begin{equation}
\phi_{M,n}(x) \equiv
\Psi_{n}(\tilde{\Omega}_{M}; x - a -
\frac{\gamma (G + 2 {\bar G} a) {\hbar}^{2} M}{m \tilde{\Omega}_{M}^{2}})
\end{equation}
where
$\tilde{\Omega}_{M}
\equiv \sqrt{\Omega^{2} - \frac{2\gamma\bar{G} \hbar M}{m}}$.

It is important to note that the squared
component of the field represented by its second derivative
$\bar{G}$ must not be too large as then it suppresses
the squared term of the particle oscillator
potential $\frac{m\Omega^{2}q^{2}}{2}$ leading to Hamiltonian
unbounded form below. This occurs for the values violating the inequality
\begin{equation}
\bar{G} < \frac{m \Omega^{2}}{2\gamma \hbar M_{max}}.
\label{diss}
\end{equation}
where $M_{max}=S$.
In such case the first term of the spectrum
(\ref{WID}) becomes imaginary for some $M$. Now, if the particle were a
light, spin endowed part of molecule
(the model which we shall discuss subsequently),
this could be interpreted as the dissociation of the molecule
caused by the strong gradient of the magnetic field
(herewith, we will keep, a bit roughly,  the term ``dissociation'').
The last inequality establishes
the boundaries of our model.
In this context it is convenient to introduce the new discrete parameter
defined by
\be
\mbar={ 2\gamma \gbar \hbar M\over \Omega^2 m}
\ee
Using $\mbar$ the inequality (\ref{diss}) writes as
\be
\mbar_{max}<1
\ee
i.e. the dissociation occurs if $\mbar$ reaches the unity.
In terms of $\mbar$, the dissociation occurs if $\mbar$ reaches unity, so that
the scope of the presented model is described by inequality $\mbar<
1$. Roughly speaking $\mbar$ is equal to  the ratio of $M$ to the number
of the maximal spin level for which  dissociation occurs. As we
are interested in the effects caused by gradients, we put $B_0=0$.
Then the energy spectrum can be written
in the following form
\bex
E_{M,n}=\hbar \Omega({1\over2} +n) \sqrt{1-\mbar}-
{m\Omega^2(G/\gbar)a \over 2}  {\mbar\over 1-\mbar} -\nonumber\\
{m\Omega^2 a^2 \over 2} {\mbar\over 1-\mbar}
- {m\Omega^2(G/\gbar)^2\over 2} {\mbar^2\over 4(1-\mbar)}
\eex

Consider first the case of infinitely small sample concentrated in
the origin of the reference frame (i.e. put $a=0$). Then we
obtain especially appealing form of energy spectrum
\be
E_{M,n}(a=0)=E^{qu}_{osc}(\Omega,n)\sqrt{1-\mbar}
-E^{cl}_{osc}(\Omega,G/\gbar){\mbar^2\over 4(1-\mbar)},
\ee
where $E^{qu}_{osc}=\hbar \Omega^2({1\over2} +n)$ and
$E^{cl}_{osc}(\Omega,G/\gbar)={ m\Omega^2(G/\gbar)^2\over 2}$.

Here  we have two competitive terms: the first one represents the energy
of {\it quantum} oscillator of frequency $\Omega$ (i.e. the energy of the particle
in absence of the field) while the second one is nothing but the
total energy of the {\it classical} oscillator of the same frequency, of the
amplitude determined by the {\it shape}  of the field.
Both terms have weights depending on the scaled spin number.
Let us now keep the value of $\gbar$ constant (suitably chosen, in order not to
be too close to the dissociation regime, e.g. to have $\mbar\simeq 0.1$).
Now varying the linear gradient  $G$ we obtain smooth transition from quantum
oscillator regime  to the classical one. Provided $G$ is sufficiently large,
so that the classical part dominates, we have a kind of {\it amplification}
of the vibrations of the particle by the magnetic field. In both cases the
frequency $\Omega$ determines the ratio of splitting of the spin levels.
Thus by measuring the latter, we can obtain  the information about the value of the
frequency. Note that, remarkably, the same is {\it impossible} in homogeneous
field: the possibility of monitoring vibrational frequency via
magnetic resonance is exclusively due to the {\it gradient} terms.

Note that the double ``quantum-classical oscillator'' structure of the energy
levels is due to quadratic gradient term.
Indeed, for $\gbar=0$ we
we cannot use the scaled spin level
number $\mbar$ as the latter is defined for $\gbar\not=0$.
In such  a case we have no "dissociation" as there
is no denominator singularity in formula (\ref{WID}).
If we put, instead, $G=0$ (with $a\not=0$ we still have two oscillators,
however the latter one follows from finite dimensions of the sample (so that
the centre of the oscillations may feel nonzero value of magnetic field)
\be
E_{M,n}(G=0)=E^{qu}_{osc}(\Omega,n)\sqrt{1-\mbar}-
E^{cl}_{osc}(\Omega,a){\mbar^2\over 1-\mbar}
\ee
On Fig. 1 we present these levels for electronic MR (EMR)
versus quadratic gradient $\gbar$ for the system of spin
$S=3/2$ (we keep $G$ being nonzero, but weak). One can observe remarkable
crossing points
for sufficiently
large $\gbar$.
Again, for a given value of $\gbar$ the level difference is
determined by the vibrational frequency $\Omega$.


Let us summarise our results. The very important feature of the
obtained spectrum of the particle is that the spin quantum number
$M$ has been {\it coupled} to the characteristic frequency $\Omega$
of the particle. This means that now for {\it fixed}
quantum oscillator number $n$ the M-dependent
energy sublevels form the structure
which {\it depends on the characteristic  frequency
$\Omega$ of the particle}.
Indeed, putting $G=\bar{G}=0$ one can immediately  see
that it {\it does not} happen for uniform
field $\mbox{\bf B} (x)=\mbox{\bf B}_{0}$.
It is remarkable result for the following reasons.
Imagine the molecules with spin-less heavy ``core''
and spin $S$ light part coupled to it.
Assume that we have a mixture of {\it unknown} molecules
of that kind, each of them possessing unknown
characteristic frequencies $\Omega_{1}$, $\Omega_{2}$, ... .
Then the fact that  the spectrum (\ref{WID})
(with nonzero $G$ or $\bar{G}$) depends on the frequency 
will provide the possibility to identify
them basing directly
on the NMR techniques that 
generically deal with $2S+1$ spin sublevels of the given level.

It is interesting that if we have $\bar{G}=0$
and nonzero $G$ then the $n$ dependent term does not depend
on quantum number $M$.
Of course, the structure of spin sublevels does not depend on $n$.
Then all the molecules of one kind
(distinguished by $\Omega$) would
give {\it the same} signal from $M$ sublevel of
{\it all} oscillatory levels.
A very interesting feature of the considered interaction is the
emerging amplification of the vibrational mode to the form of classical
oscillator. The effect is due to nonzero $\gbar$.
The amplitude is determined by the shape of the field, and the
amplification is obtained for high values of the linear gradient.

Concerning the experimental realiation of the presented spin-oscillator
coupling effect note that for the typical molecular frequences ($\approx
10^{14} Hz$) one would need enormously strong gradients to observe the
effect. Indeed, the strongest gradients obtained so far amount to $10^6 T/m$
\cite{Sidles92}. Now, for molecule with  incorporated atom of oxygene
the above
gradients allow to realize the effect for frequences up to $10^4 Hz$.
Besides the highest frequency for which the effect can be observed satisfies
roughly $\overline{G}\sim \Omega^2$. On the other   hand, there exist molecular
systems that exhibit low frequences \cite{low}. Thus the effect
is difficult but still possible to realize experimentally.

In conclusion, we have demonstrated that MR spectroscopy with parabolic
magnetic field provides  subtle   spin-oscillator coupling  effects, that
allow to select the characteristic frequency of the system. It seems to open
new  prospects for the MR applications such as  quantum level separations
\cite{Lewa96a,Lewa96b} or quantum computation \cite{NMRQC}.
Finally, we expect that the idea of incorporating higher order gradients will
prove fruitful in looking  for new solutions within MR domain.
We also believe that the obtained results will  stimulate
development of techniques of producing strong gradient field of different shapes.

We are grateful to Prof. Zdzis\l{}aw Paj{\c a}k for very helpful discussion.
Cz. L. was supported by the Polish Committee for Scientific Research, contract
No. 2P03B 14812. M. H., P. H. and R.H. were partially supported by
Polish Committee for Scientific Research, Contract No. 2 P03B 103 16 and
the European Science Foundation. M.H. and P.H. also
acknowledge the support from the Foundation for Polish Science.

\begin{figure}
\label{fig1}
\caption[ll]{Structure of the EMR energy levels (a) in parabolic field
for finite dimension of the sample ($S=3/2$, $a=10^{-4}m$, $\Omega=10^5Hz$,
$G=-0.003T/m$) and oscillatory spectrum (b).
}
\end{figure}
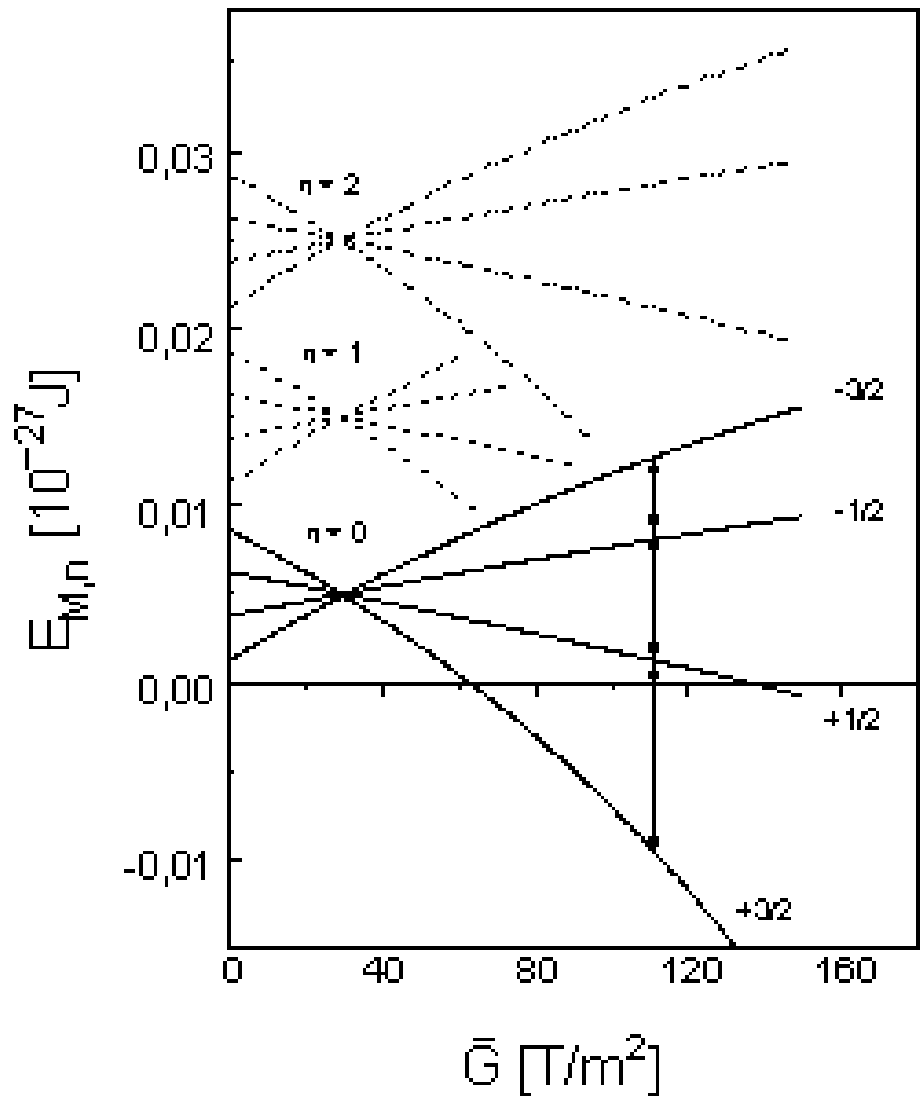
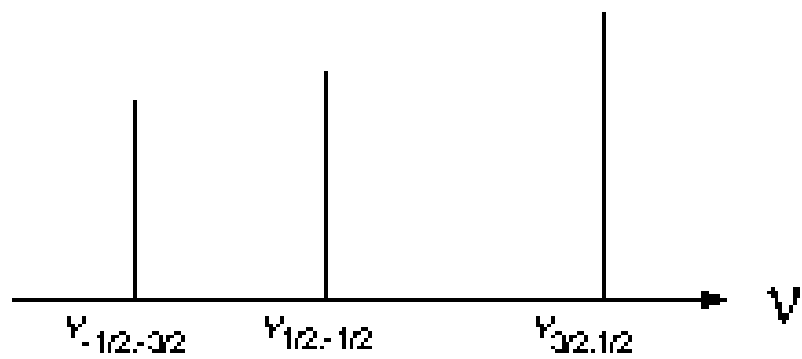
\end{document}